\documentclass[twocolumn,preprintnumbers,superscriptaddress,nofootinbib,aps,prd,floatfix]{revtex4}

\usepackage{amsmath,amssymb}
\usepackage{graphicx}
\usepackage{subfigure}
\usepackage{hyperref}
\usepackage{footmisc}

\usepackage{array}
\newcolumntype{L}[1]{>{\raggedright\let\newline\\\arraybackslash\hspace{0pt}}m{#1}}
\newcolumntype{C}[1]{>{\centering\let\newline\\\arraybackslash\hspace{0pt}}m{#1}}
\newcolumntype{R}[1]{>{\raggedleft\let\newline\\\arraybackslash\hspace{0pt}}m{#1}}

\newcommand{\be}{\begin{eqnarray*}}
\newcommand{\ee}{\end{eqnarray*}}

\newcommand{\bee}{\begin{eqnarray}}
\newcommand{\eee}{\end{eqnarray}}
\newcommand{\beeq}{\begin{equation}}
\newcommand{\eeeq}{\end{equation}}

\newcommand{\ep}{\varepsilon}

\newcommand{\zzf}{\Pi_{ZZ}(m_Z^2)}
\newcommand{\zzfn}{\Pi_{ZZ}(0)}
\newcommand{\azf}{\Pi_{\gamma Z}(m_Z^2)}
\newcommand{\azfn}{\Pi_{\gamma Z}(0)}
\newcommand{\aaf}{\Pi_{\gamma \gamma}(m_Z^2)}

\newcommand{\wwf}{\Pi_{WW}(m_W^2)}
\newcommand{\wwfn}{\Pi_{WW}(0)}
\newcommand{\alfa}{\alpha}

\begin{document}

\title{Disformal dark energy at colliders}
\begin{abstract}
  Disformally coupled, light scalar fields arise in many of the
  theories of dark energy and modified gravity that attempt to explain
  the accelerated expansion of the universe. They have proved
  difficult to constrain with precision tests of gravity because they
  do not give rise to fifth forces around static non-relativistic
  sources. However, because the scalar field couples derivatively to
  standard model matter, measurements at high energy particle
  colliders offer an effective way to constrain and potentially detect
  a disformally coupled scalar field. Here we derive new constraints
  on the strength of the disformal coupling from LHC run 1 data and
  provide a forecast for the improvement of these constraints from run
  2. We additionally comment on the running of disformal and standard model
  couplings in this scenario under the renormalisation group flow.
\end{abstract}
%
%

\author{Philippe Brax} 
\email{Philippe.Brax@cea.fr}
\affiliation{Institut de Physique Th\'{e}orique, Universit\'e Paris-Saclay, CEA, CNRS, F-91191 Gif/Yvette Cedex, France\\[0.1cm]}

\author{Clare Burrage} 
\email{Clare.Burrage@nottingham.ac.uk}
\affiliation{School of Physics and Astronomy, University of Nottingham, Nottingham, NG7 2RD, United Kingdom\\[0.1cm]}

\author{Christoph Englert} 
\email{Christoph.Englert@glasgow.ac.uk}
\affiliation{SUPA, School of Physics and Astronomy, University of Glasgow, Glasgow, G12 8QQ, United Kingdom\\[0.1cm]}

\pacs{}
\preprint{}
\maketitle

\section{Introduction}
\label{sec:intro}

Evidence for the acceleration of the expansion of the universe comes
from a wide variety of cosmological
observations~\cite{Efstathiou:1990xe,Perlmutter:1998np,Riess:1998cb,Lahav:2014vza},
which probe the expansion history of the universe and the way in which
the distribution of light and matter has evolved to form structures.
There is currently no convincing theoretical explanation for this
expansion.  Introduction of a cosmological constant requires an
extreme fine tuning to explain why its value is so small, when quantum
fluctuations of standard model fields want to drive its value to the
highest energy scale of the theory~\cite{Weinberg:1988cp}. In
contrast to the hierarchy problem, the cosmological constant problem
is a fine tuning problem that exists even at low energies (smaller
than the electroweak scale) within the Standard Model. Attempts to
solve the cosmological constant problem, either by introducing new
fields or by modifying the gravitational
sector~\cite{Copeland:2006wr,Clifton:2011jh}, typically suffer from
either related fine-tuning problems or an inability to match current
observations. Almost all attempts to solve the cosmological constant
problem introduce new, light scalar degrees of freedom that we will
call dark energy.\footnote{We use the term dark energy to include any
  scalar field that is introduced as part of a solution to the
  cosmological constant problem, not just those that directly drive
  the expansion of the universe to accelerate.} Therefore, even
without knowing the complete explanation for the accelerated expansion
of the universe, we can learn about the form of the underlying theory
by studying the behaviour of the resulting dark energy scalars.

As dark energy is part of a hypothesised solution to the cosmological
constant problem it is expected to interact with both standard model
and gravitational fields~\cite{Joyce:2014kja}. The expectation that
dark energy would couple to standard model fields has proved
particularly difficult to embed in the theory, as Yukawa type
interactions are excluded by the results of fifth force
searches~\cite{Adelberger:2003zx} to a high degree of accuracy.
However these measurements only restrict one particular class of
interaction between dark energy and the standard model, those that
arise through a conformal coupling where matter fields move on
geodesics of a metric $\tilde{g}_{\mu\nu} = A(\phi)g_{\mu\nu}$, where
$g_{\mu\nu}$ is the spacetime metric and $A$ an arbitrary function of
the dark energy scalar field. A second class of interactions, termed
disformal, is possible. In a disformal theory matter fields move on
geodesics of the metric $\tilde{g}_{\mu\nu}= g_{\mu\nu}
+B(\phi) \partial_{\mu}\phi\partial_{\nu} \phi$, where again $B$ is an
arbitrary function of $\phi$. Disformal interactions have been shown
to arise in the four dimensional effective theory resulting from
various brane world scenarios~\cite{deRham:2010eu,Koivisto:2013fta},
in branon models~\cite{Alcaraz:2002iu,Cembranos:2004jp} and in
theories of massive
gravity~\cite{deRham:2010kj,deRham:2010ik}. Disformal couplings are
particularly interesting in theories where an (approximate) shift
symmetry for the scalar field is used to protect the mass of the dark
energy scalar, and ensure that it can remain light on cosmological
scales. Unlike conformal couplings, disformal couplings to matter do
not break this shift symmetry. One prime example is provided by the
Goldstone modes of a global symmetry where the interaction potential
results from a soft and explicit breaking of the symmetry. Axion
quintessence models fall into this category and are an example of a
thawing model of dark energy \cite{Frieman:1995pm}. These theories do
not make a definitive prediction for the scale of the disformal
interaction, allowing it to lie anywhere between the dark energy scale
$\Lambda \sim 10^{-3} \mbox{ eV}$ and the Planck scale $M_P \sim
10^{18} \mbox{ GeV}$. It must therefore be determined by experiment.

In contrast to conformal couplings, which are tightly constrained by
experiments, disformal couplings have proved difficult to study
experimentally. In particular disformal interactions hide from fifth
force searches extremely successfully because they are not sourced by
static, non-relativistic matter distributions. A new approach is
needed to study disformal dark energy interactions.

Disformal interactions are derivative interactions between a light
scalar field and matter, therefore they can be most efficiently probed
at high energies. The possibility of using particle colliders to
constrain disformally coupled scalars was first proposed by Kaloper
\cite{Kaloper:2003yf}. In previous work, two of us have shown that
such couplings can be studied and constrained both in terrestrial
laboratories and from observations of stars~\cite{Brax:2014vva}. A
constraint was estimated from early mono-photon searches at the Large
Hadron Collider (LHC) and required the coupling scale $M \gtrsim 10^2
\mbox{ GeV}$. A comparable constraint was obtained from restricting
additional energy losses in supernovae. These constraints are eleven
orders of magnitude stronger than the bounds that can be obtained from
local tests of gravity~\cite{Sakstein:2014isa}. A variety of other
observational probes of disformal couplings have been previously
considered: the disformal interactions of scalars with photons can be
probed in laboratory experiments~\cite{Brax:2012ie} and astrophysical
observations~\cite{Brax:2015fya}. In models motivated by Galileon
theories and massive gravity, constraints have been put on the
disformal interactions from studying gravitational lensing and the
velocity dispersion of galaxies~\cite{Wyman:2011mp,Sjors:2011iv}.
Further cosmological implications of disformal scalars have been
considered in
\cite{Zumalacarregui:2010wj,Koivisto:2012za,Bettoni:2012xv,vandeBruck:2013yxa,Brax:2013nsa,Neveu:2014vua,vandeBruck:2015ida,Brax:2014vla}.

We briefly review the
models considered in this paper in section
\ref{sec:dsca}. In section~\ref{sec:elwp} we consider constraints on this class of models
that arise from precision measurements at the Large Electron Positron
Collider (LEP). Specifically, we compute the
oblique corrections and investigate the impact of disformally coupled
scalars on the $Z$ boson line shape. Section~\ref{sec:LHC} is devoted
to setting constraints from LHC measurements using the full 
run 1 data sets available from both ATLAS and CMS. In particular we
study constraints from di-lepton, mono-photon and mono-jet production
in association with missing transverse energy and extrapolate
promising channels to the end of run 2. We also comment on
modifications to the recently discovered Higgs boson phenomenology
that arise from disformally coupled scalars. Throughout we will use a ``mostly minus'' convention
for the Minkowski metric.

\section{Disformal Dark Energy}
\label{sec:dsca}
A scalar field couples disformally to matter if matter fields move on
geodesics of the metric~\cite{Bekenstein:1992pj}
\begin{equation}
\tilde{g}_{\mu\nu}= g_{\mu\nu} + B(\phi) \partial_{\mu}\phi \partial_{\nu}\phi\,,
\label{eq:metric}
\end{equation}
where $g_{\mu\nu}$ is the space time metric, and $B$ is an arbitrary
function of the scalar field $\phi$. In this work we will only
consider $B(\phi)= 1/M^4$, where $M$ is a constant with dimensions of
mass. This is the leading order term in a Taylor expansion of
$B(\phi)$ and will be sufficient to demonstrate the effects of a
disformal coupling at the LHC which will appear first at order
$(E/M)^2$, where $E$ is the characteristic energy of the process under
discussion. The value of $\phi$ may very from place to place in the
universe leading to a possible redressing of the scale $M$, which
would need to be taken into account in order to compare constraints
derived from different places correctly. We will not discuss
variations in the background value of $\phi$ further in this work.

 Matter fields are conserved with
respect to the metric in equation (\ref{eq:metric}) so that
\begin{equation}
\tilde{D}_{\mu} \tilde{T}^{\mu\nu}=0\,,
\end{equation}
where $\tilde{D}_{\mu}$ is the covariant derivative with respect to
the disformal metric of equation (\ref{eq:metric}), and
\begin{equation}
\tilde{T}^{\mu\nu} = \frac{2}{\sqrt{-\tilde{g}}}\frac{\delta S_m}{\delta \tilde{g}_{\mu\nu}}
\end{equation}
is the Jordan frame energy momentum tensor.

The interactions between the scalar field $\phi$ and standard model
fields that arise from interactions with the metric in equation
(\ref{eq:metric}) occur at all orders in $(E/M)^4$, however if we are
only interested in the leading order effects of the disformal coupling
then the relevant interaction terms in the action are
\begin{equation}
  \label{eq:disf}
  S = \frac{1}{M^4} \int d^4 x \sqrt{g} \partial_{\mu} \phi \partial _{\nu}\phi T^{\mu\nu}\,,
\end{equation}
where now $T^{\mu\nu} = (2/\sqrt{-g})(\delta S_m/\delta g_{\mu\nu})$
is the Einstein frame energy-momentum tensor, defined with respect to
the space-time metric $g_{\mu\nu}$. In this work we restrict our
attention to interactions of the form given in Equation
(\ref{eq:disf}), other possible derivative interactions of the same
mass dimension are discussed in the Appendix. However we will show in
Section \ref{sec:running} that if they are initially chosen to be
absent additional operators are not generated by quantum corrections
at order $1/M^4$.

Equation~\eqref{eq:disf} motivates two approaches to observe or
constrain models with disformally coupled scalars at colliders. The
first avenue is through modifications to the standard model
expectation for processes at precision machines such as LEP through
either direct production of the scalar, which modifies for example the
$Z$ boson phenomenology or through internal quantum corrections that
lead to a deviation from the standard model (SM) expectation, for
example for two to two particle scattering. The second avenue is
provided by exploiting the current energy frontier of the
LHC. Derivative disformal couplings induce deviations at large
momentum transfers, which are directly accessible to current
searches. Since the scalar $\phi$ is light and stable on collider
scales, established dark matter searches for missing
energy~\cite{cmsjets,cmsphoton,atlasphoton,atlasdilepton} also provide
sensitive strategies to look for disformal couplings.
\begin{figure}[b!]
  \includegraphics[width=0.23\textwidth]{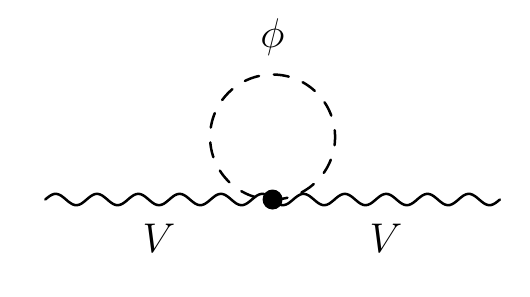}
  \vspace{-0.3cm}
  \caption{\label{fig:polarization} One-loop and leading $1/M^4$
    contribution to the vector $V=\gamma,W^\pm,Z$ polarization
    functions $\Pi_{\mu\nu}(q^2)=\Pi(q^2)g_{\mu\nu} + \dots$ mediated
    by a virtual disformal scalar coupling.}
\end{figure}

\section{Constraints from electroweak precision measurements}
\label{sec:elwp}
\subsection{Oblique corrections}
\label{sec:oblique}
A customary way of assessing the impact of new physics on standard
model processes is via its impact on precision measurements performed
during the LEP era. A framework which is typically adopted to analyse
modifications in the gauge sector are the oblique corrections
parametrised by the $S,T,U$ parameters of Peskin and
Takeuchi~\cite{Peskin:1991sw,Peskin:1990zt}:
\begin{subequations}
  \label{eq:peskintak}
  \begin{widetext}
    \begin{align}  
      S&= {4 s_w^2c_w^2 \over \alfa} \left(\frac{\zzf - \zzfn}{m_Z^2}
        - {c_w^2-s_w^2\over c_ws_w}{\azf-\azfn\over m_Z^2} - {\aaf\over m_Z^2}\right)\,,\\
      T&= {1\over \alfa}\left( {\wwfn\over m_W^2} - {\zzfn\over m_Z^2}
        - {2 s_w\over c_w} {\azfn\over m_z^2}\right) \,, \\
      U&= {4 s_w^2\over \alfa} \left( {\wwf-\wwfn\over m_W^2} - c_w^2
        {\zzf - \zzfn\over m_Z^2 } \right.
      - \left. s_w^2{\aaf\over m_Z^2} -2 s_w c_w
        {\azf-\azfn\over m_Z^2} \right)\,,
    \end{align}
  \end{widetext}
\end{subequations}
\begin{figure*}[!t]
  \centering
  \parbox{0.45\textwidth}{\vspace{-6cm}\includegraphics[height=0.18\textwidth]{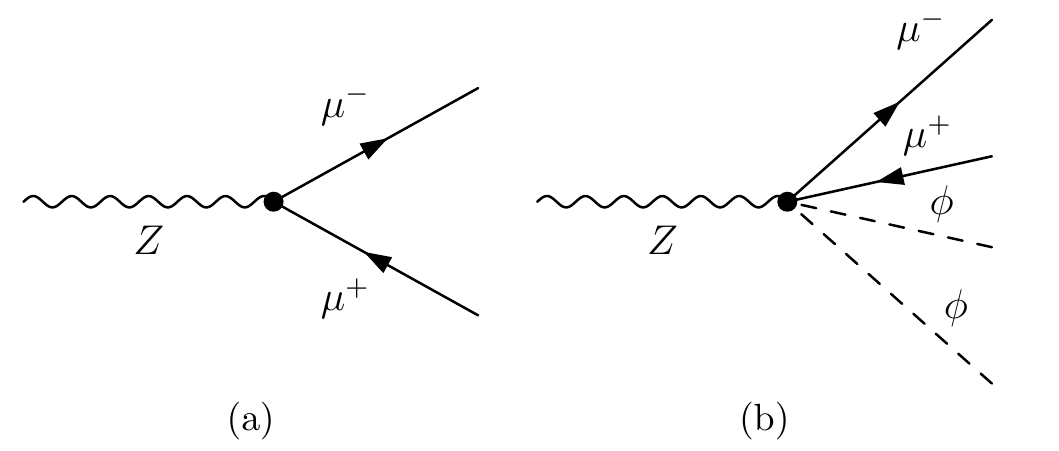}}\qquad
  \includegraphics[height=0.3\textwidth]{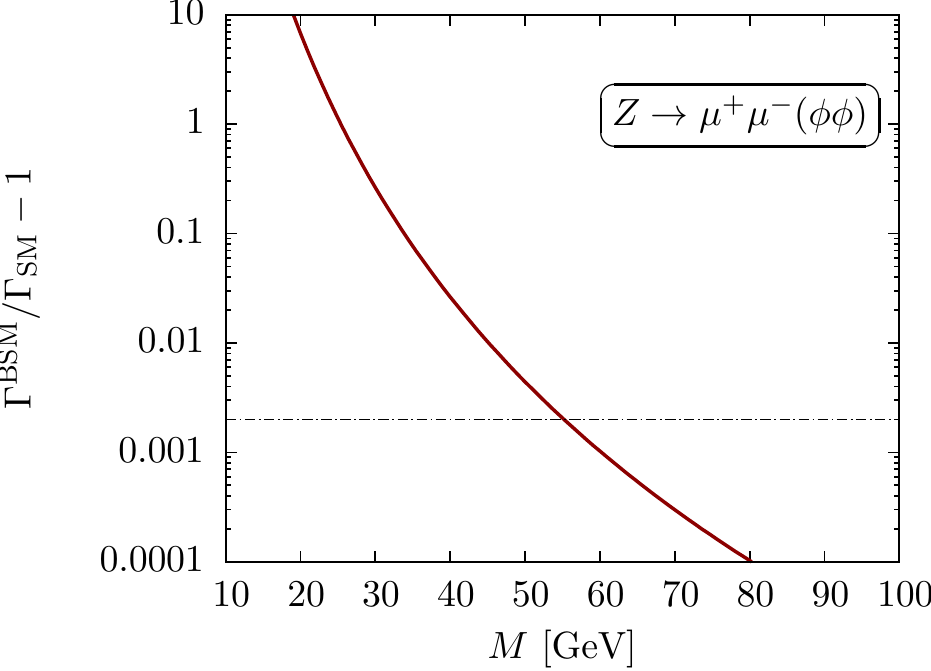}
  \caption{\label{fig:zdec} Modification of the $Z$ boson width to
    leptons (here concretely for $Z\to \mu^+\mu^-$) due to
    ``dressing'' the decay using the new interaction with two
    scalars. The leading order Feynman diagram contributing to this
    decay is shown in (a) and an example diagram contributing to the
    modification of the decay at $\sim 1/M^4$ is shown in (b). Note
    that the individual fermion legs and the $Z$ boson propagator can
    also be dressed with a $\phi^2$ insertion and these diagrams are
    not shown. Similarly the production of the $Z$ boson receives
    modifications. The vertical line represents the current bound on
    $Z\to \mu^+\mu^-$~\cite{Agashe:2014kda}.}
\end{figure*}
where the $\Pi$ functions denote SM vector boson polarization
functions and $c_W,s_W$ are cosine and sine of the weak mixing angle,
respectively. From their definition it becomes obvious that the
Peskin-Takeuchi parameters capture beyond the standard model
(BSM)-induced effects in the gauge sector in a $q^2$ expansion of the
polarization functions to leading order. $S,T,U$ parametrise
``universal'' modifications due to BSM physics in the gauge sector,
i.e. the parameters provide an approximation to the full
next-to-leading order results under the assumption that the BSM
physics arises in the gauge sector only. As such, consistency with
existing constraints on the Peskin-Takeuchi parameters should not be
understood as consistency of a particular model with electroweak
precision measurements, it merely acts as a first test that a
particular model has to pass.

The contributions of the disformally coupled scalar at one loop and
leading order in $1/M^4$ are sketched in
Fig.~\ref{fig:polarization}. The polarization function for the photon
reads:
\begin{alignat}{5}
  \Pi_{\gamma\gamma}(q^2) &= {q^2\over 32 \pi^2}
  \left({m_\phi\over M}\right)^4\,.
\end{alignat} 
This is consistent with intact
gauge invariance. 
For broken directions (and $Z-\gamma$ mixing) we have
\begin{alignat}{4}
  \Pi_{\gamma Z} (q^2) &= 0 \,,\\
  \Pi_{VV} (q^2) &= {1\over 128\pi^2}\, \left( m_\phi^2 ( 4q^2-3m_V^2)
    \right. \nonumber \\ & \hspace{1cm} \left. + 10m_V^2
    A_0(m_\phi^2) \right) \left( {m_\phi \over M^2} \right)^2 \,,
\end{alignat} 
where $V=W^\pm, Z$ and $A_0(m^2)$ is the scalar loop function in the
Passarino Veltman language~\cite{Passarino:1978jh,Denner:1991kt} in
$D$-dimensional regularisation
\begin{multline}
A_0(m^2) = {(2\pi \mu)^{4-D}\over i\pi^2} \int {\hbox{d}^D q } \,
{1\over q^2 - m^2}  \\
= -m^2 \left( m^2 \over 4\pi\mu^2\right)^{D/2 - 2}
  \Gamma\left(1-{D\over 2}\right)\,,
\end{multline}
where $\mu$ is the so-called `t Hooft mass that keeps track of mass
units in $D$ dimensions and cancels in renormalised quantities. With
these equations, it is easy to see that the contributions of the $\sim
q^0,q^2$ pieces to the Peskin-Takeuchi parameters
Eq.~\eqref{eq:peskintak} vanish identically. Note that due to the
appearance of only $A_0$ functions in the gauge boson self-energies
$\sim q^2$ there are no contributions to the extended set of precision
observables as defined in~\cite{Barbieri:2004qk}, which capture the
impact of BSM effects on the vector boson self-energies $\sim
q^4$. Therefore no constraints can be placed on the disformal coupling
scale $M$ from precision measurements of $S, T$ and $U$. We will
comment on the impact on running couplings in
section~\ref{sec:running}.

\subsection{$Z$ boson phenomenology}
Another important and precisely determined quantity is the lineshape
of the $Z$ boson. Since the disformal coupling dresses every
interaction vertex and the scalar mass can be significantly below the
$Z$ boson threshold a novel $1\to 4$ channel will open at leading
order in the $1/M^4$ expansion. Depending on the size of $M$, this can
lead to a significant modification of the $Z$ bosons decay
phenomenology and the $Z$ boson lineshape as a consequence. The size
of the BSM correction as a function of $M$ is shown in
Fig.~\ref{fig:zdec}. We can use this to set a lower limit on $M$;
however an explicit calculation shows that this lower bound is only
weak, $M\gtrsim 60~\text{GeV}$.

\section{Constraints from LHC searches}
\label{sec:LHC}
The form of the disformal coupling means that disformal scalars can be
produced on shell in a collider and, as they will not interact in the
detector, will leave only missing energy as a signature of their
presence. This means that missing energy searches for dark matter can
be adapted to place constraints on disformally coupled dark energy
models. In this section we recast recent analyses performed by ATLAS
and CMS in the mono-jet~\cite{cmsjets}, mono-photon
\cite{cmsphoton,atlasphoton} and di-lepton~\cite{atlasdilepton}
searches, where available, to reinterpret these measurements and set
constraints on the scale $M$.\footnote{Mono-lepton
  searches~\cite{cmslepton,atlaslepton} crucially depend on a correct
  modeling of the missing energy resolution, and we do not consider
  these searches.} By validating our analysis against the 8 TeV
results we can also extrapolate our findings to the upcoming run 2 and
estimate the limit that will be set in the near future.

For the actual analyses we include the signal and all dominant
backgrounds and simulate them with {\sc{FeynRules}}~\cite{feynrules},
{\sc{MadEvent}}~\cite{madevent}, {\sc{Sherpa}}~\cite{sherpa} and
{\sc{Herwig++}}~\cite{herwig}, jet clustering is performed with
{\sc{FastJet}}~\cite{fastjet}. Throughout the numerical analysis we
make the specific choice that $m_\phi=1~\text{MeV}$. However our
results will be independent of the mass of such a light scalar, and so
will be valid for all lighter masses.

Setting limits on effective field theories at colliders can suffer
from shortcomings if the new physics scale is resolved by the
experiment~\cite{Englert:2014cva,Buchmueller:2013dya,Harris:2014hga,Jacques:2015zha,Buckley:2014fba,Haisch:2015ioa}. An
effective field theory description is only valid if the probed energy
scales are lower than the scale of new physics, e.g. if resonances or
thresholds remain unresolved. These shortcomings can be mended by
reverting to concrete UV complete scenarios (but limits become model
dependent as a consequence) or by separating energy scales
consistently on the basis of renormalisation group equations
\cite{Englert:2014cva,Isidori:2013cla}. In the scenario we consider in
this paper it is important to highlight a difference compared to
similar issues in dark matter searches: while complete field theoretic
models can be constructed in dark matter-related analyses, a concrete
model implementation is not available in the analysis of strongly
coupled gravitational effects due to the intrinsic non-linear and
non-renormalisable nature of gravity in a perturbative field theory
context. In what follows, this issue should be kept in mind; while
limit setting is a viable qualitative strategy in the absence of a
signal, the interpretation of a possible excess seen with $M$ in the
TeV region will require the inclusion of non-linear effects which are
formally higher order in our $(E/M)^4$ expansion. We will comment on
the validity of the set limits in light of resolved energy scales
later in section~\ref{sec:running}.

\subsubsection*{Mono-Photon searches}
Both ATLAS and CMS have published analyses in mono-photon searches for
the full run 1 data set~\cite{cmsphoton,atlasphoton} with similar
sensitivities.

CMS reconstruct jets using the anti-$k_T$ algorithm~\cite{antikt} to
cluster particles into jets with resolution parameter\footnote{For readers less
  familiar with jet physics the $D$ parameter refers to the conical
  size of the jet combined from particle tracks in the azimuthal
  angle--pseudo-rapidity plane. An excellent review of jet physics is
  provided in \cite{Salam:2009jx}.} $D=0.5$ and define isolated
photons based on the energy deposit in a cone of $\Delta R
=\sqrt{\Delta\Phi^2+\Delta \eta^2} < 0.3$ (where $\Phi$ and $\eta$ are the 
azimuthal angle and pseudo-rapidity, respectively) around the photon
candidate, which is required to be smaller than 5\% of the candidate's
energy based on the expected shower profile. The transverse energy of
the photon is required to be $E_T(\gamma)>145~\text{GeV}$. Since a
full detector simulation is not available, we approximate this energy
with the Monte Carlo information for the transverse momentum. Events
with more than one jet with $p_T>30~\text{GeV}$ and light leptons
(isolation is based on a hadronic energy deposit in the vicinity of
$\Delta R <0.3$ by less than 20\% of the candidate $p_T$) with
$p_T>10~\text{GeV}$ are vetoed if they are separated from the photon
by $\Delta R>0.5$. The final selection requires a missing transverse
energy ${{E}}_T^{\text{miss}}>140~\text{GeV}$, well separated from the
photon in azimuthal angle $\Delta \Phi(E_T^{\text{miss}},\gamma)>2$. We
include an expected missing energy resolution by fitting the
expectation of CMS particle flow as outlined in
\cite{Englert:2012wf}. After these steps CMS exclude an upper cross
section limit of 14~fb, which translates into a lower limit of
\begin{alignat}{5}
\hbox{CMS:} & \quad & M\gtrsim 419~\text{GeV}
\end{alignat}
in our analysis.

ATLAS follow a similar strategy, selecting photons with
$p_T(\gamma)>125~\text{GeV}$ in $|\eta_\gamma|<1.37$,
$E_T^{\text{miss}}>150~\text{GeV}$ and $\Delta\Phi
(E_T^{\text{miss}},\gamma)>0.4$. Jets are reconstructed with the
anti-$k_T$ algorithm with $D=0.4$ and vetoed if
$p_{T,j}>30~\text{GeV}$ and $\Delta\Phi
(E_T^{\text{miss}},j)<0.4$. Electrons ($p_T>7~\text{GeV}$,
$|\eta|<2.47$) and muons ($p_T>6~\text{GeV}$, $|\eta|<2.5$) are
vetoed. ATLAS exclude 6.1 events at 95\% confidence level for the run
1 luminosity of 20.1/fb. This translates in our implementation into
\begin{alignat}{5}
  \hbox{ATLAS:} & \quad & M\gtrsim 447~\text{GeV}\,, 
\end{alignat}
which is consistent with the CMS limit.

\begin{figure}[!t]
  \centering  
  \includegraphics[height=0.3\textwidth]{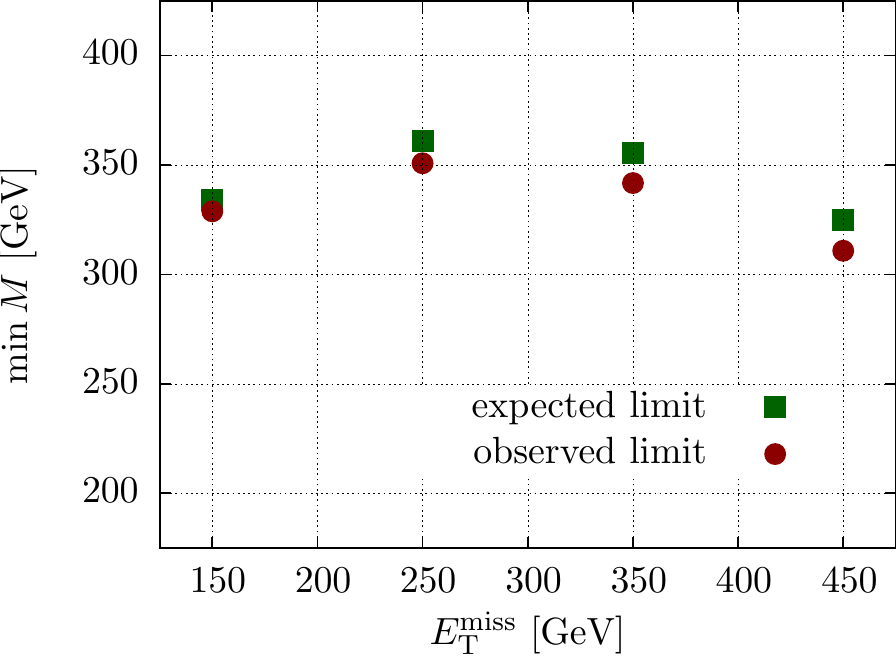}
  \caption{\label{fig:atlaslepton} Results of the ATLAS dilepton
    search confronted with disformal scalar coupling model. For
    further details see the text. ATLAS do not quote uncertainties and
    hence we only show central values.}
\end{figure}

\subsubsection*{Di-Lepton searches}
ATLAS have published a dark matter search in a $Z+$missing energy
search based on the run 1 data set in~\cite{atlasdilepton}. In their
analysis, ATLAS require electrons to have $E_T>20~\text{GeV},
|\eta|<2.47$ and consider muons with $p_T>20~\text{GeV},
|\eta|<2.5$. Isolation is defined by requiring the hadronic energy
deposit in a cone of size $\Delta R<0.2$ around the candidate to be
less than 10\% of the candidate's $E_T$ and only tracks with
$p_T>1~\text{GeV}$ are considered in this isolation criterion. Jets
are reconstructed with the anti-$k_T$ algorithm with $D=0.4$, and
$p_T>25~\text{GeV}$ and $|\eta|<2.5$.

Candidate events need to have a di-lepton system consistent with the
$Z$ boson $76~\text{GeV}\leq m_{\ell \ell} \leq 106~\text{GeV}$ and the
missing energy has to be well separated from the di-lepton pair: $\Delta
\Phi (E_T^{\text{miss}},\ell\ell) >2.5$. Further, ATLAS impose $\eta^{\ell
  \ell}< 2.5$, $|p_T^{\ell\ell} - E_T^{\text{miss}}|/p_T^{\ell\ell}<0.5$. Events
with jets with $p_T> 25~\text{GeV}$ are finally removed and ATLAS
consider four search regions based on a inclusive selection of
missing energy and provide expected and observed fiducial cross
sections at 95\% confidence level.

Implementing these analysis steps, we can again translate these limits
into lower limits on the disformal coupling scale $M$, as depicted in
Fig.~\ref{fig:atlaslepton}. The sensitivity is maximised for the
$E_T^{\text{miss}}>250~\text{GeV}$ search region, where the trade off
between differential signal cross section enhancement due to the
probed energy in the disformal coupling and decreasing background
cross sections becomes optimal. For more stringent requirements, the
signal becomes kinematically suppressed.

\subsubsection*{Mono-Jet searches}
The most recent mono-jet analysis exploiting the full run 1 data set
has been provided by the CMS collaboration in~\cite{cmsjets}. In this
analysis particles are clustered into jets using the anti-$k_T$ algorithm
\cite{antikt} with $D=0.5$ and requiring the leading jet
to have transverse momentum and rapidity
\begin{equation}
  p_{T,j_1}>110~\text{GeV}, |\eta_{j_1}| < 2.4\,.
\end{equation}
A second jet
\begin{equation}
  p_{T,j_2}>30~\text{GeV}, |\eta_{j_2}| < 4.5
\end{equation}
is allowed if it is separated from the first jet by
\begin{equation}
  \Delta\Phi(j_1,j_2) < 2.5\,.
\end{equation}
The analysis vetos events with more than two jets with
$p_T>30~\text{GeV}$ and $|\eta_j|<4.5$. Events with isolated leptons
are vetoed if $p_{T,\ell}>10~\text{GeV}$; isolation is defined by
requiring the total hadronic energy deposit in a cone of size $0.4$
around the lepton candidate being smaller than 20\% of its transverse
momentum. The analysis selects 7 inclusive search regions based on an
additional missing energy threshold.

\begin{figure}[!t]
  \centering
  \includegraphics[height=0.3\textwidth]{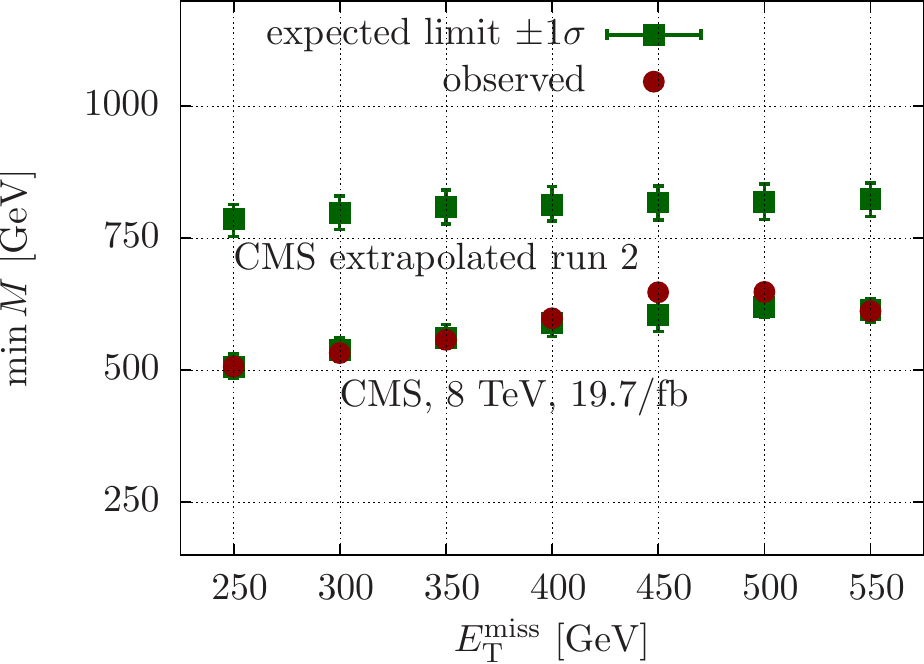}
  \caption{\label{fig:cmsjet} Minimum scale $M$ extracted from the CMS
    monojet search of~\cite{cmsjets} in the different search regions
    based on the inclusive search regions characterised by
    $E_{T}^{\text{miss}}$ (for details see text). We also show the
    improvement based on an extrapolation of the 8 TeV analysis to the
    13 TeV LHC run 2 with 100/fb.}
\end{figure}

We validate our analysis in these different search regions, taking
into account the $Z+\text{jets}$, QCD-jets, $t\bar t+\text{jets}$,
$W+\text{jets}$ and diboson+jets backgrounds and missing energy
resolution is included in an analogous manner to that of the previous sections. We find
excellent agreement of our analysis with the CMS background templates,
especially for the dominant backgrounds and most inclusive
selections. The agreement of our simulation with the $Z+\text{jets}$
template in particular, provides confidence that we can set a
trustable limit on the presence of an additional missing energy
contribution that is the main signature of our model in this
channel. For the fake-dominated background contributions as QCD jets,
we use the CMS results to derive a differential efficiency, which we
will later use in our 13 TeV projections without further modification.

The results of the 8 TeV CMS analysis, recast along the above lines is
shown in Fig.~\ref{fig:cmsjet}. Comparing the findings of the mono-jet
search to the previously discussed channels, we see that the mono-jet
search is the most sensitive to our scenario. In light of this result
we extrapolate the 8 TeV CMS to LHC run 2 in Fig.~\ref{fig:cmsjet}
using the CLs method of~\cite{cls1,cls2}. Most of this improvement
stems from a significant signal cross section increase by a factor of
10. It is not entirely unexpected that this particular analysis
performs better than the other channels discussed above. The
particular form of the interaction basically amounts to QCD-like BSM
production suppressed by the scale $M$, and the relative sensitivity
follows the paradigm that events induced by strong couplings give
tighter constraints on new physics than those produced by weak
couplings because jet production is the most abundant high transverse
momentum process at the LHC.

\subsubsection*{A note on modified Higgs phenomenology}
\begin{figure}[!t]
  \centering
  \includegraphics[height=0.3\textwidth]{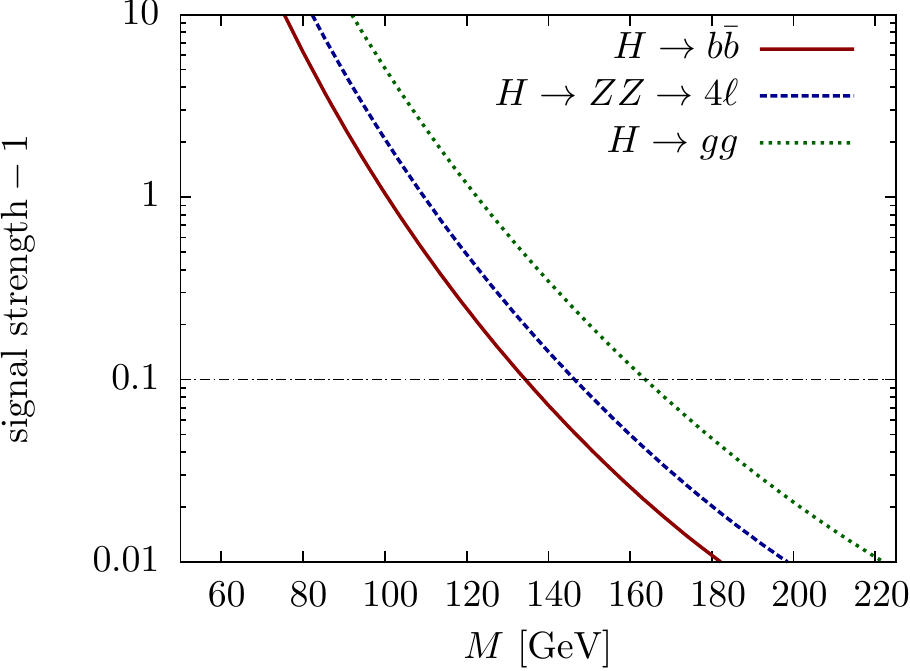}
  \caption{\label{fig:modh} Modifications of the Higgs boson signal
    strength as a function of $M$ for the gluon fusion production mode
    at 8 TeV, estimated using Higgs effective field
    theory~\cite{Kniehl:1995tn,Ellis:1975ap,Shifman:1979eb}. The
    modifications are analogous to the 2-body and 4-body decay sample
    Feynman diagrams shown in Fig.~\ref{fig:zdec}.}
\end{figure}

Finally we comment on potential modification of Higgs
phenomenology. The crucial observable is the ``signal strength'' 
\begin{equation}
  \mu_{ik}={[\sigma_{\{i \}} (H) \times \text{BR}(H\to \{ k\} )]^{\text{BSM}} \over
    [\sigma_{\{i\}} (H) \times \text{BR}(H\to \{ k \} )]^{\text{SM}}}\,,
\end{equation}
this measures the cross sections for Higgs production via mechanism
$i$ and subsequent decay into final state $k$ with decay probability
$\text{BR}(h\to \{ k\} )$, relative to the SM expectation of the same
production and decay. Measurements of the Higgs signal strength have
already reached considerable sensitivity $\sim 10\%$ around the SM
hypothesis, but are mostly driven by the gluon fusion production
mechanism. It is expected that we can scrutinize the Higgs boson's
phenomenology at the percent level at the high luminosity
LHC~\cite{Englert:2014uua,Klute:2013cx}. ``Dressing'' the Higgs
vertices with the additional interactions analogous to the $Z$ boson
phenomenology (see Fig.~\ref{fig:zdec}) at LEP, we can understand the
allowed error as a limit on the scale $M$ with the benefit of a higher
mass scale $m_h>m_Z$. The modifications of the signal strength for the
different decay modes are shown in Fig.~\ref{fig:modh}. Whilst Higgs
phenomenology is sensitive to the presence of a disformal scalar,
direct searches for missing energy remain a more powerful probe.

\section{Running of couplings}
\label{sec:running}
In the previous sections we have focused on direct measurements of and constraints on the
presence of a disformal interaction. In section \ref{sec:oblique} we saw
that the disformal scalar does not lead to oblique
corrections. Not all BSM effects are expressed through oblique
corrections and we analyse other observables in the following. We will
see that these effects are highly suppressed and limited to
kinematic thresholds in the limit $m_\phi/M\ll 1$, where we can trust
our expansion in terms of an effective field theory deformation of the
SM. We will also comment on potential modifications of the running of
couplings due to the presence of the disformal interactions.

\begin{figure*}[!t]
  \centering
  \includegraphics[width=0.75\textwidth]{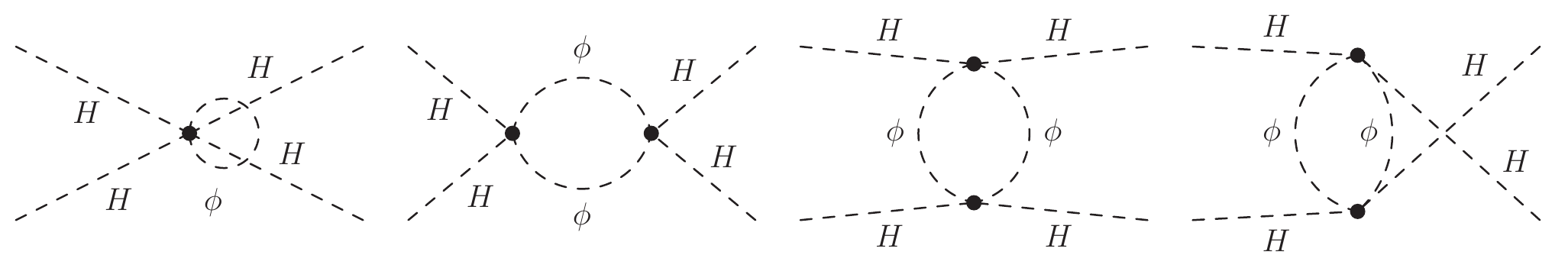}
  \caption{\label{fig:h4} One-loop corrections to the $H^4$
    operator at order $1/M^4$.}
\end{figure*}

\begin{figure*}[!t]
  \centering
  \includegraphics[width=0.75\textwidth]{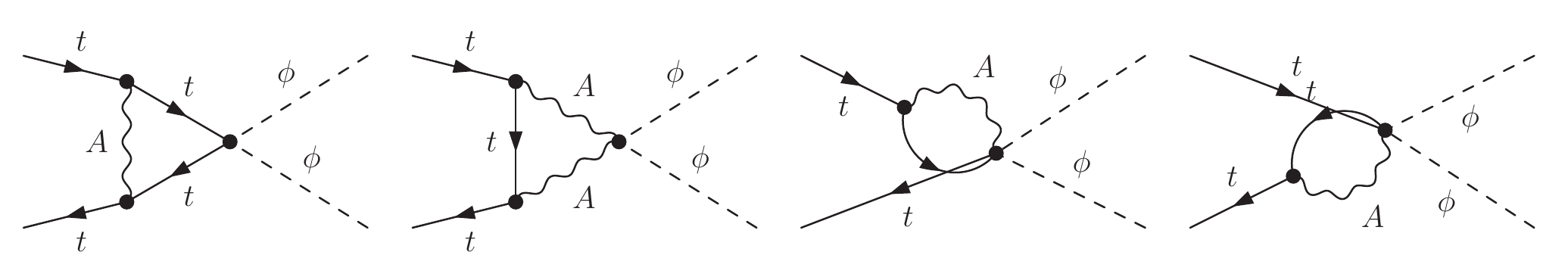}
  \caption{\label{fig:ttphiphi} One-loop corrections to the $t\bar t
    \phi \phi$ vertex in the considered toy model, extended by the
    disformal coupling.}
\end{figure*}

Taking the contribution to the vector polarization functions as a
starting point we can check whether the presence of the additional
derivatively-coupled scalars impacts the running of SM couplings. In
particular any impact on the top Yukawa couplings, as well as on the
Higgs self interaction will have important consequences on the
stability of the electroweak vacuum. By working out the $\sim M^{-4}$
corrections to wave function renormalisations and SM vertices we can
compute potential contributions to the SM $\beta$
functions. Introducing the $\overline{\hbox{MS}}$ parameter in
$D=4-2\ep$ dimensional regularisation 
\begin{equation}
  \Delta = { \Gamma(1+\ep)\over \ep } \left({4\pi\mu^2\over \mu_R^2}\right)^\ep
\end{equation}
we have $A_0(m^2)=m^2\Delta + {\cal{O}}(\ep)$. $\mu_R$ is the
renormalisation scale that effectively replaces $\mu$ in the
renormalisation procedure. We can calculate
the renormalisation $\sim 1/M^4$ of the Higgs and top wave functions
in the $\overline{\hbox{MS}}$ scheme (note that all corrections vanish
in the limit $m_\phi\to 0$ in dimensional regularisation)
\begin{alignat}{5}
  \delta Z_H &=&{1 \over 32\pi^2}\, \Delta \,\left(m_\phi\over M\right)^4 \,,\\
  \delta Z_t&=&{3 \over 64\pi^2} \, \Delta \, \left(m_\phi\over M\right)^4\,.
\end{alignat}
The renormalisation of the $\bar t tH  $ vertex to $\sim
1/M^4$ is
\begin{equation}
   \delta Z_{\bar t t H }= {1 \over 16\pi^2} \,\, \Delta \, \left(m_\phi\over M\right)^4\,,
\end{equation}
which leads to an operator renormalisation 
\begin{equation}
  \delta Z = \delta Z_{\bar t t H } - \delta Z_t - {1\over 2} \delta Z_H = 0\,,
\end{equation}
which implies no contribution $\sim 1/M^4$ to the anomalous dimension
of the Yukawa coupling $y_t$ and its running is therefore not affected by the
presence of the disformal coupling.

Similarly we can compute the renormalisation of the quartic
interaction operator in the Higgs potential by investigating the four
point vertex $H^4$
(Fig.~\ref{fig:h4})
\begin{equation}
  \delta Z_{H^4}={3 \over 8\pi^2} \, \Delta \, \left(m_\phi\over M\right)^4 
\end{equation}
and again we find an operator renormalisation 
\begin{equation}
  \delta Z = \delta Z_{H^4} - 2 \, \delta Z_H = 0\,,
\end{equation}
meaning that again the disformal scalar has no impact on the running
of the quartic Higgs interaction.

We have checked that similar cancellations happen in the
renormalisation of all other SM couplings. This cancellation is not an
accident, but happens due to the independence of the external and
internal symmetries of the considered effective field
theory. 

The deformation of the disformal interactions, on the other hand, has
a dynamic dependence on scale which, if external and internal
symmetries are independent, should be limited to the interaction
itself as well as invariants of the external (diffeomorphisim)
symmetry. Due to the complicated dynamics of the SM as a whole, to
analyse this we focus on a simple subsector of the SM; QED with the
top as a single massive fermion. Concretely we look at the
renormalisation of the operator
\begin{equation}
  \label{eq:model}
  {\cal{O}}_{\text{dis}}={c_T\over M^4} T^{\mu\nu}\partial_\mu\phi \partial_{\nu} \phi
\end{equation}
and specifically at the dressed top propagator as probe of the Wilson
coefficient $c_T$.

In this model, the wave function renormalisation for the scalar $\phi$
is given by
\begin{equation}
  \delta Z_\phi = {21\over 8\pi^2} \, \Delta \, \left(m_t\over M\right)^4
\end{equation}
and the full top renormalisation, including the $1/M^0$ part in
general gauge is
\begin{equation}
  \delta Z_t = {\alpha \over 9 \pi} \, \Delta \, \xi +  {3\over
    64\pi^2} \, \Delta \, \left( m_\phi \over M\right)^4\,,
\end{equation}
where $\xi$ is the gauge parameter. The $\phi^2$-dressed top
propagator, Fig.~\ref{fig:ttphiphi} is renormalised by
\begin{equation}
  \label{eq:ttphiphi}
  \delta Z_{t\bar t \phi^2}= {\alpha \over 9 \pi} \, \Delta \, \xi
\end{equation}
and therefore the operator renormalisation constant is~(see
e.g. \cite{Buchalla:1995vs} for a detailed discussion of
renormalisation in effective field theories)
\begin{equation}
  \label{eq:opren}
  \delta Z_{{\cal{O}}_{\text{dis}}} = -{3\over 64\pi^2} \, \Delta \, \left( {m_\phi \over M}
  \right)^4  - {21 \over 8\pi^2} \, \Delta \, \left( {m_t \over M}
  \right)^4\,,
\end{equation}
which implies an anomalous dimension
\begin{multline}
  \gamma_{c_T}= {\hbox{d}\, \delta Z_{{\cal{O}}_{\text{dis}}} \over \hbox{d}\log \mu_R} =
  {1\over 16\pi^2}\left[ {3\over 2} \, \left( {m_\phi \over M}
    \right)^4 + 84 \left( {m_t \over M} \right)^4 \right]\,,
\end{multline}
such that the renormalisation group equation (RGE) reads
\begin{equation}
  \label{eq:limitset}
c_T(\Lambda) = \left( {\Lambda
      \over M }\right)^{\gamma_{c_T}}\quad (\Lambda\leq M)
\end{equation}
after inserting the boundary condition $c_T(M)=1$. Since the anomalous
dimension is positive definite, it flows to small coupling in the
infrared, consistent with the behavior of a coupling that parametrises
the interaction with SM energy-momentum. It is important to note that
when calculating the operator renormalisation in Eq.~\eqref{eq:opren}
we do not obtain spurious singularities, which would need to be
absorbed by additional counter terms unrelated to
${\cal{O}}_{\text{dis}}$. The presence of such terms would be
tantamount to a RGE flow-induced mixing of $\cal{O}_{\text{dis}}$ with
other independent operators\footnote{Independent in this context means
  that redundancies are removed with equations of motion.} that are
excluded from our effective theory of Eq.~\eqref{eq:model} at the UV
scale (by construction). While such terms could in principle be
present, see the appendix, our constraints can be understood as
consistent limits on the operator ${\cal{O}}_{\text{dis}}$ alone.

Furthermore, in the previous sections we have found limits in the 650
GeV range, which are mass scales easily resolved by the LHC at 8
TeV. In principle this raises the question of whether we can trust our
effective field theory prescription if the most sensitive region to
the presence of disformal couplings is given by $p_{T,j}>M$ in, e.g.,
the mono-jet analysis. However, equipped with the above RGE equations
we can separate the scale of measurement and new physics consistently
(see e.g. \cite{Englert:2014cva} for a related discussion in Higgs
phenomenology): If the new physics scale is indeed higher than the
resolved scale we can compute the modified limit using RGE equations
like Eq.~\eqref{eq:limitset}. Since the top quark is the heaviest
particle in the SM, we can expect that Eq.~\eqref{eq:limitset}
also gives a reasonable estimate of the size of these effects in the
full SM. If we push the fundamental scale $M$ outside the LHC coverage
the effectively resolved scale due to Eq.~\eqref{eq:limitset}
remains numerically unchanged due to the smallness of the anomalous
dimension $\gamma_{c_T}$ and our limit is solid against these
aforementioned issues.

The findings for the running of the disformal coupling deserve a few
additional remarks. Firstly, the renormalisation is not only
gauge-invariant, i.e. the terms $\sim \xi$ drop out, but all terms
$\sim 1/M^0$ cancel in the calculation. This means that the internal
symmetries which are expressed as the zeroth order in the $\sim
M^{-4}$ expansion do not influence the running of the disformal
coupling. The leading term in the renormalisation only depends on
invariants of Lorentz symmetry (masses and gauge couplings) and arises
purely from wave function renormalisation constants
(Eq.~\eqref{eq:ttphiphi} is pure gauge), meaning that the coupling
remains universal under renormalisation group flow. This is an
explicit realisation of the result derived in~\cite{Hui:2010dn} by Hui
and Nicolis, who demonstrated that once a universal coupling between a
scalar field and matter has been postulated, this coupling is stable
against classical and quantum renormalisations in the matter sector.

If we take the calculation at face value (and neglect the potential
presence of higher order terms) the running of $c_T$ (with UV boundary
condition $c_T(M)=1$) is given as a function of all masses in the
theory, which displays the running of the energy momentum tensor by
all explicit sources of the breaking of conformal
invariance.\footnote{It is worth pointing out that this result
  therefore crucially relies on dimensional regularisation to avoid
  spurious terms~\cite{Englert:2013gz}. } The running is not
influenced by the gauge couplings, this demonstrates that the dynamics
of internal and external symmetries factorize reminiscent of the general
structure discussed by Coleman and Mandula~\cite{Coleman:1967ad}.

\section{Conclusions}
If dark energy couples to matter disformally our best chances to
detect these interactions come from events occurring at high energies:
While precision measurements at LEP provide a bound on $M$, the LHC
offers the best current prospects for such a study, and we have shown
that mono-jet searches performed by the CMS collaboration provide the
best current constraint on the energy scale of the disformal coupling
$M\gtrsim 650~\text{GeV}$. The particular form of the interactions,
coupling to the energy momentum tensor, decouples the disformal scalar
from precision electroweak observables as well as from the running of
SM couplings, in agreement with the expectation of a factorisation of
outer and inner symmetries in interacting QFT. This leaves direct
detection as the main collider avenue to set constraints on the
presence of disformal couplings. To this end we extrapolate the CMS
mono-jet result to the higher energy collisions at the 13 TeV LHC run
2 and estimate that the 8 TeV result can be improved to $M\gtrsim
750~\text{GeV}$ for a run 2 luminosity of $100~\text{fb}^{-1}$, or
even higher if systematics are improved.

\acknowledgments We would like to thank Nemanja Kaloper and David
Seery for very helpful discussions during the preparation of this
work. PhB acknowledges partial support from the European Union FP7 ITN
INVISIBLES (Marie Curie Actions, PITN- GA-2011- 289442) and from the
Agence Nationale de la Recherche under contract ANR 2010 BLANC 0413
01. CB is supported by a Royal Society University Research
Fellowship. CE is supported in part by the IPPP Associateship
programme and is grateful to the Mainz Institute for Theoretical
Physics (MITP) for its hospitality and its partial support during the
completion of this work.

\appendix*
\section{Additional interactions}
\label{app}
In this work we have studied the leading order behaviour when matter
fields move on geodesics of a purely disformal metric. However if we
wanted to relax this assumption there are a two other possible types
of operator that would allow the scalar field to interact with matter
in a universal way that have the same mass dimension as those
considered here. Firstly, the scalar field may also couple conformally
to matter. If this coupling is purely a function of the scalar
derivatives then the leading order interaction with matter is:
\begin{equation}
  (\beta/M^4)(\partial \phi)^2 T
\label{eq:betaT}
\end{equation}
for constant $\beta$. Secondly, additional terms arise because the
energy momentum tensor of the matter fields is only uniquely defined
up to addition of terms proportional to the equations of motion. In
general relativity the energy momentum tensor remains finite under the
renormalisation group flow after allowing for the inclusion of such
terms \cite{Brown:1992db}. For the Higgs scalar in Minkowski space,
for example, this allows for the inclusion the term
\begin{equation}
  R(\tilde{g}) \vert H\vert^2
\end{equation}
where $\tilde{g}_{\mu\nu}$ is the disformal metric of
Eq.~(\ref{eq:metric}), and $R$ is the associated Ricci scalar.
Written explicitly in terms of the disformal scalar this allows for
the inclusion of interactions of the form
\begin{equation}
  \frac{\gamma}{M^4}((\Box \phi)^2 - \nabla_\mu \nabla_\nu \phi \nabla^\mu \nabla^\nu \phi) \vert H\vert^2
\label{eq:Rphi}
\end{equation}
We stress, however, that if $\beta$ and $\gamma$ are assumed initially
to be zero, as in the main body of this article, then they remain zero
under the renormalisation group flow, at least to order $(E/M)^4$. If
these terms are allowed to be non-zero then the renormalisation group
is expected to mix these coefficients with $c_T$ via the equations of
motion.

\bibliographystyle{apsrev}
\bibliography{draft}

\begin{thebibliography}{71}
\expandafter\ifx\csname natexlab\endcsname\relax\def\natexlab#1{#1}\fi
\expandafter\ifx\csname bibnamefont\endcsname\relax
  \def\bibnamefont#1{#1}\fi
\expandafter\ifx\csname bibfnamefont\endcsname\relax
  \def\bibfnamefont#1{#1}\fi
\expandafter\ifx\csname citenamefont\endcsname\relax
  \def\citenamefont#1{#1}\fi
\expandafter\ifx\csname url\endcsname\relax
  \def\url#1{\texttt{#1}}\fi
\expandafter\ifx\csname urlprefix\endcsname\relax\def\urlprefix{URL }\fi
\providecommand{\bibinfo}[2]{#2}
\providecommand{\eprint}[2][]{\url{#2}}

\bibitem[{\citenamefont{Efstathiou et~al.}(1990)\citenamefont{Efstathiou,
  Sutherland, and Maddox}}]{Efstathiou:1990xe}
\bibinfo{author}{\bibfnamefont{G.}~\bibnamefont{Efstathiou}},
  \bibinfo{author}{\bibfnamefont{W.}~\bibnamefont{Sutherland}},
  \bibnamefont{and} \bibinfo{author}{\bibfnamefont{S.}~\bibnamefont{Maddox}},
  \bibinfo{journal}{Nature} \textbf{\bibinfo{volume}{348}},
  \bibinfo{pages}{705} (\bibinfo{year}{1990}).

\bibitem[{\citenamefont{Perlmutter et~al.}(1999)}]{Perlmutter:1998np}
\bibinfo{author}{\bibfnamefont{S.}~\bibnamefont{Perlmutter}}
  \bibnamefont{et~al.} (\bibinfo{collaboration}{Supernova Cosmology Project}),
  \bibinfo{journal}{Astrophys.J.} \textbf{\bibinfo{volume}{517}},
  \bibinfo{pages}{565} (\bibinfo{year}{1999}), \eprint{astro-ph/9812133}.

\bibitem[{\citenamefont{Riess et~al.}(1998)}]{Riess:1998cb}
\bibinfo{author}{\bibfnamefont{A.~G.} \bibnamefont{Riess}} \bibnamefont{et~al.}
  (\bibinfo{collaboration}{Supernova Search Team}),
  \bibinfo{journal}{Astron.J.} \textbf{\bibinfo{volume}{116}},
  \bibinfo{pages}{1009} (\bibinfo{year}{1998}), \eprint{astro-ph/9805201}.

\bibitem[{\citenamefont{Lahav and Liddle}(2014)}]{Lahav:2014vza}
\bibinfo{author}{\bibfnamefont{O.}~\bibnamefont{Lahav}} \bibnamefont{and}
  \bibinfo{author}{\bibfnamefont{A.~R.} \bibnamefont{Liddle}}
  (\bibinfo{year}{2014}), \eprint{1401.1389}.

\bibitem[{\citenamefont{Weinberg}(1989)}]{Weinberg:1988cp}
\bibinfo{author}{\bibfnamefont{S.}~\bibnamefont{Weinberg}},
  \bibinfo{journal}{Rev.Mod.Phys.} \textbf{\bibinfo{volume}{61}},
  \bibinfo{pages}{1} (\bibinfo{year}{1989}).

\bibitem[{\citenamefont{Copeland et~al.}(2006)\citenamefont{Copeland, Sami, and
  Tsujikawa}}]{Copeland:2006wr}
\bibinfo{author}{\bibfnamefont{E.~J.} \bibnamefont{Copeland}},
  \bibinfo{author}{\bibfnamefont{M.}~\bibnamefont{Sami}}, \bibnamefont{and}
  \bibinfo{author}{\bibfnamefont{S.}~\bibnamefont{Tsujikawa}},
  \bibinfo{journal}{Int.J.Mod.Phys.} \textbf{\bibinfo{volume}{D15}},
  \bibinfo{pages}{1753} (\bibinfo{year}{2006}), \eprint{hep-th/0603057}.

\bibitem[{\citenamefont{Clifton et~al.}(2012)\citenamefont{Clifton, Ferreira,
  Padilla, and Skordis}}]{Clifton:2011jh}
\bibinfo{author}{\bibfnamefont{T.}~\bibnamefont{Clifton}},
  \bibinfo{author}{\bibfnamefont{P.~G.} \bibnamefont{Ferreira}},
  \bibinfo{author}{\bibfnamefont{A.}~\bibnamefont{Padilla}}, \bibnamefont{and}
  \bibinfo{author}{\bibfnamefont{C.}~\bibnamefont{Skordis}},
  \bibinfo{journal}{Phys.Rept.} \textbf{\bibinfo{volume}{513}},
  \bibinfo{pages}{1} (\bibinfo{year}{2012}), \eprint{1106.2476}.

\bibitem[{\citenamefont{Joyce et~al.}(2015)\citenamefont{Joyce, Jain, Khoury,
  and Trodden}}]{Joyce:2014kja}
\bibinfo{author}{\bibfnamefont{A.}~\bibnamefont{Joyce}},
  \bibinfo{author}{\bibfnamefont{B.}~\bibnamefont{Jain}},
  \bibinfo{author}{\bibfnamefont{J.}~\bibnamefont{Khoury}}, \bibnamefont{and}
  \bibinfo{author}{\bibfnamefont{M.}~\bibnamefont{Trodden}},
  \bibinfo{journal}{Phys.Rept.} \textbf{\bibinfo{volume}{568}},
  \bibinfo{pages}{1} (\bibinfo{year}{2015}), \eprint{1407.0059}.

\bibitem[{\citenamefont{Adelberger et~al.}(2003)\citenamefont{Adelberger,
  Heckel, and Nelson}}]{Adelberger:2003zx}
\bibinfo{author}{\bibfnamefont{E.}~\bibnamefont{Adelberger}},
  \bibinfo{author}{\bibfnamefont{B.~R.} \bibnamefont{Heckel}},
  \bibnamefont{and} \bibinfo{author}{\bibfnamefont{A.}~\bibnamefont{Nelson}},
  \bibinfo{journal}{Ann.Rev.Nucl.Part.Sci.} \textbf{\bibinfo{volume}{53}},
  \bibinfo{pages}{77} (\bibinfo{year}{2003}), \eprint{hep-ph/0307284}.

\bibitem[{\citenamefont{de~Rham and Tolley}(2010)}]{deRham:2010eu}
\bibinfo{author}{\bibfnamefont{C.}~\bibnamefont{de~Rham}} \bibnamefont{and}
  \bibinfo{author}{\bibfnamefont{A.~J.} \bibnamefont{Tolley}},
  \bibinfo{journal}{JCAP} \textbf{\bibinfo{volume}{1005}}, \bibinfo{pages}{015}
  (\bibinfo{year}{2010}), \eprint{1003.5917}.

\bibitem[{\citenamefont{Koivisto et~al.}(2014)\citenamefont{Koivisto, Wills,
  and Zavala}}]{Koivisto:2013fta}
\bibinfo{author}{\bibfnamefont{T.}~\bibnamefont{Koivisto}},
  \bibinfo{author}{\bibfnamefont{D.}~\bibnamefont{Wills}}, \bibnamefont{and}
  \bibinfo{author}{\bibfnamefont{I.}~\bibnamefont{Zavala}},
  \bibinfo{journal}{JCAP} \textbf{\bibinfo{volume}{1406}}, \bibinfo{pages}{036}
  (\bibinfo{year}{2014}), \eprint{1312.2597}.

\bibitem[{\citenamefont{Alcaraz et~al.}(2003)\citenamefont{Alcaraz, Cembranos,
  Dobado, and Maroto}}]{Alcaraz:2002iu}
\bibinfo{author}{\bibfnamefont{J.}~\bibnamefont{Alcaraz}},
  \bibinfo{author}{\bibfnamefont{J.}~\bibnamefont{Cembranos}},
  \bibinfo{author}{\bibfnamefont{A.}~\bibnamefont{Dobado}}, \bibnamefont{and}
  \bibinfo{author}{\bibfnamefont{A.~L.} \bibnamefont{Maroto}},
  \bibinfo{journal}{Phys.Rev.} \textbf{\bibinfo{volume}{D67}},
  \bibinfo{pages}{075010} (\bibinfo{year}{2003}), \eprint{hep-ph/0212269}.

\bibitem[{\citenamefont{Cembranos et~al.}(2004)\citenamefont{Cembranos, Dobado,
  and Maroto}}]{Cembranos:2004jp}
\bibinfo{author}{\bibfnamefont{J.}~\bibnamefont{Cembranos}},
  \bibinfo{author}{\bibfnamefont{A.}~\bibnamefont{Dobado}}, \bibnamefont{and}
  \bibinfo{author}{\bibfnamefont{A.~L.} \bibnamefont{Maroto}},
  \bibinfo{journal}{Phys.Rev.} \textbf{\bibinfo{volume}{D70}},
  \bibinfo{pages}{096001} (\bibinfo{year}{2004}), \eprint{hep-ph/0405286}.

\bibitem[{\citenamefont{de~Rham et~al.}(2011)\citenamefont{de~Rham, Gabadadze,
  and Tolley}}]{deRham:2010kj}
\bibinfo{author}{\bibfnamefont{C.}~\bibnamefont{de~Rham}},
  \bibinfo{author}{\bibfnamefont{G.}~\bibnamefont{Gabadadze}},
  \bibnamefont{and} \bibinfo{author}{\bibfnamefont{A.~J.}
  \bibnamefont{Tolley}}, \bibinfo{journal}{Phys.Rev.Lett.}
  \textbf{\bibinfo{volume}{106}}, \bibinfo{pages}{231101}
  (\bibinfo{year}{2011}), \eprint{1011.1232}.

\bibitem[{\citenamefont{de~Rham and Gabadadze}(2010)}]{deRham:2010ik}
\bibinfo{author}{\bibfnamefont{C.}~\bibnamefont{de~Rham}} \bibnamefont{and}
  \bibinfo{author}{\bibfnamefont{G.}~\bibnamefont{Gabadadze}},
  \bibinfo{journal}{Phys.Rev.} \textbf{\bibinfo{volume}{D82}},
  \bibinfo{pages}{044020} (\bibinfo{year}{2010}), \eprint{1007.0443}.

\bibitem[{\citenamefont{Frieman et~al.}(1995)\citenamefont{Frieman, Hill,
  Stebbins, and Waga}}]{Frieman:1995pm}
\bibinfo{author}{\bibfnamefont{J.~A.} \bibnamefont{Frieman}},
  \bibinfo{author}{\bibfnamefont{C.~T.} \bibnamefont{Hill}},
  \bibinfo{author}{\bibfnamefont{A.}~\bibnamefont{Stebbins}}, \bibnamefont{and}
  \bibinfo{author}{\bibfnamefont{I.}~\bibnamefont{Waga}},
  \bibinfo{journal}{Phys.Rev.Lett.} \textbf{\bibinfo{volume}{75}},
  \bibinfo{pages}{2077} (\bibinfo{year}{1995}), \eprint{astro-ph/9505060}.

\bibitem[{\citenamefont{Kaloper}(2004)}]{Kaloper:2003yf}
\bibinfo{author}{\bibfnamefont{N.}~\bibnamefont{Kaloper}},
  \bibinfo{journal}{Phys.Lett.} \textbf{\bibinfo{volume}{B583}},
  \bibinfo{pages}{1} (\bibinfo{year}{2004}), \eprint{hep-ph/0312002}.

\bibitem[{\citenamefont{Brax and Burrage}(2014)}]{Brax:2014vva}
\bibinfo{author}{\bibfnamefont{P.}~\bibnamefont{Brax}} \bibnamefont{and}
  \bibinfo{author}{\bibfnamefont{C.}~\bibnamefont{Burrage}},
  \bibinfo{journal}{Phys.Rev.} \textbf{\bibinfo{volume}{D90}},
  \bibinfo{pages}{104009} (\bibinfo{year}{2014}), \eprint{1407.1861}.

\bibitem[{\citenamefont{Sakstein}(2014)}]{Sakstein:2014isa}
\bibinfo{author}{\bibfnamefont{J.}~\bibnamefont{Sakstein}},
  \bibinfo{journal}{JCAP} \textbf{\bibinfo{volume}{1412}}, \bibinfo{pages}{012}
  (\bibinfo{year}{2014}), \eprint{1409.1734}.

\bibitem[{\citenamefont{Brax et~al.}(2012)\citenamefont{Brax, Burrage, and
  Davis}}]{Brax:2012ie}
\bibinfo{author}{\bibfnamefont{P.}~\bibnamefont{Brax}},
  \bibinfo{author}{\bibfnamefont{C.}~\bibnamefont{Burrage}}, \bibnamefont{and}
  \bibinfo{author}{\bibfnamefont{A.-C.} \bibnamefont{Davis}},
  \bibinfo{journal}{JCAP} \textbf{\bibinfo{volume}{1210}}, \bibinfo{pages}{016}
  (\bibinfo{year}{2012}), \eprint{1206.1809}.

\bibitem[{\citenamefont{Brax et~al.}(2015{\natexlab{a}})\citenamefont{Brax,
  Brun, and Wouters}}]{Brax:2015fya}
\bibinfo{author}{\bibfnamefont{P.}~\bibnamefont{Brax}},
  \bibinfo{author}{\bibfnamefont{P.}~\bibnamefont{Brun}}, \bibnamefont{and}
  \bibinfo{author}{\bibfnamefont{D.}~\bibnamefont{Wouters}}
  (\bibinfo{year}{2015}{\natexlab{a}}), \eprint{1505.01020}.

\bibitem[{\citenamefont{Wyman}(2011)}]{Wyman:2011mp}
\bibinfo{author}{\bibfnamefont{M.}~\bibnamefont{Wyman}},
  \bibinfo{journal}{Phys.Rev.Lett.} \textbf{\bibinfo{volume}{106}},
  \bibinfo{pages}{201102} (\bibinfo{year}{2011}), \eprint{1101.1295}.

\bibitem[{\citenamefont{Sjors and Mortsell}(2013)}]{Sjors:2011iv}
\bibinfo{author}{\bibfnamefont{S.}~\bibnamefont{Sjors}} \bibnamefont{and}
  \bibinfo{author}{\bibfnamefont{E.}~\bibnamefont{Mortsell}},
  \bibinfo{journal}{JHEP} \textbf{\bibinfo{volume}{1302}}, \bibinfo{pages}{080}
  (\bibinfo{year}{2013}), \eprint{1111.5961}.

\bibitem[{\citenamefont{Zumalacarregui
  et~al.}(2010)\citenamefont{Zumalacarregui, Koivisto, Mota, and
  Ruiz-Lapuente}}]{Zumalacarregui:2010wj}
\bibinfo{author}{\bibfnamefont{M.}~\bibnamefont{Zumalacarregui}},
  \bibinfo{author}{\bibfnamefont{T.}~\bibnamefont{Koivisto}},
  \bibinfo{author}{\bibfnamefont{D.}~\bibnamefont{Mota}}, \bibnamefont{and}
  \bibinfo{author}{\bibfnamefont{P.}~\bibnamefont{Ruiz-Lapuente}},
  \bibinfo{journal}{JCAP} \textbf{\bibinfo{volume}{1005}}, \bibinfo{pages}{038}
  (\bibinfo{year}{2010}), \eprint{1004.2684}.

\bibitem[{\citenamefont{Koivisto et~al.}(2012)\citenamefont{Koivisto, Mota, and
  Zumalacarregui}}]{Koivisto:2012za}
\bibinfo{author}{\bibfnamefont{T.~S.} \bibnamefont{Koivisto}},
  \bibinfo{author}{\bibfnamefont{D.~F.} \bibnamefont{Mota}}, \bibnamefont{and}
  \bibinfo{author}{\bibfnamefont{M.}~\bibnamefont{Zumalacarregui}},
  \bibinfo{journal}{Phys.Rev.Lett.} \textbf{\bibinfo{volume}{109}},
  \bibinfo{pages}{241102} (\bibinfo{year}{2012}), \eprint{1205.3167}.

\bibitem[{\citenamefont{Bettoni et~al.}(2012)\citenamefont{Bettoni, Pettorino,
  Liberati, and Baccigalupi}}]{Bettoni:2012xv}
\bibinfo{author}{\bibfnamefont{D.}~\bibnamefont{Bettoni}},
  \bibinfo{author}{\bibfnamefont{V.}~\bibnamefont{Pettorino}},
  \bibinfo{author}{\bibfnamefont{S.}~\bibnamefont{Liberati}}, \bibnamefont{and}
  \bibinfo{author}{\bibfnamefont{C.}~\bibnamefont{Baccigalupi}},
  \bibinfo{journal}{JCAP} \textbf{\bibinfo{volume}{1207}}, \bibinfo{pages}{027}
  (\bibinfo{year}{2012}), \eprint{1203.5735}.

\bibitem[{\citenamefont{van~de Bruck et~al.}(2013)\citenamefont{van~de Bruck,
  Morrice, and Vu}}]{vandeBruck:2013yxa}
\bibinfo{author}{\bibfnamefont{C.}~\bibnamefont{van~de Bruck}},
  \bibinfo{author}{\bibfnamefont{J.}~\bibnamefont{Morrice}}, \bibnamefont{and}
  \bibinfo{author}{\bibfnamefont{S.}~\bibnamefont{Vu}},
  \bibinfo{journal}{Phys.Rev.Lett.} \textbf{\bibinfo{volume}{111}},
  \bibinfo{pages}{161302} (\bibinfo{year}{2013}), \eprint{1303.1773}.

\bibitem[{\citenamefont{Brax et~al.}(2013)\citenamefont{Brax, Burrage, Davis,
  and Gubitosi}}]{Brax:2013nsa}
\bibinfo{author}{\bibfnamefont{P.}~\bibnamefont{Brax}},
  \bibinfo{author}{\bibfnamefont{C.}~\bibnamefont{Burrage}},
  \bibinfo{author}{\bibfnamefont{A.-C.} \bibnamefont{Davis}}, \bibnamefont{and}
  \bibinfo{author}{\bibfnamefont{G.}~\bibnamefont{Gubitosi}},
  \bibinfo{journal}{JCAP} \textbf{\bibinfo{volume}{1311}}, \bibinfo{pages}{001}
  (\bibinfo{year}{2013}), \eprint{1306.4168}.

\bibitem[{\citenamefont{Neveu et~al.}(2014)\citenamefont{Neveu,
  Ruhlmann-Kleider, Astier, Besançon, Conley et~al.}}]{Neveu:2014vua}
\bibinfo{author}{\bibfnamefont{J.}~\bibnamefont{Neveu}},
  \bibinfo{author}{\bibfnamefont{V.}~\bibnamefont{Ruhlmann-Kleider}},
  \bibinfo{author}{\bibfnamefont{P.}~\bibnamefont{Astier}},
  \bibinfo{author}{\bibfnamefont{M.}~\bibnamefont{Besançon}},
  \bibinfo{author}{\bibfnamefont{A.}~\bibnamefont{Conley}},
  \bibnamefont{et~al.}, \bibinfo{journal}{Astron.Astrophys.}
  \textbf{\bibinfo{volume}{569}}, \bibinfo{pages}{A90} (\bibinfo{year}{2014}),
  \eprint{1403.0854}.

\bibitem[{\citenamefont{van~de Bruck and Morrice}(2015)}]{vandeBruck:2015ida}
\bibinfo{author}{\bibfnamefont{C.}~\bibnamefont{van~de Bruck}}
  \bibnamefont{and} \bibinfo{author}{\bibfnamefont{J.}~\bibnamefont{Morrice}}
  (\bibinfo{year}{2015}), \eprint{1501.03073}.

\bibitem[{\citenamefont{Brax et~al.}(2015{\natexlab{b}})\citenamefont{Brax,
  Burrage, Davis, and Gubitosi}}]{Brax:2014vla}
\bibinfo{author}{\bibfnamefont{P.}~\bibnamefont{Brax}},
  \bibinfo{author}{\bibfnamefont{C.}~\bibnamefont{Burrage}},
  \bibinfo{author}{\bibfnamefont{A.-C.} \bibnamefont{Davis}}, \bibnamefont{and}
  \bibinfo{author}{\bibfnamefont{G.}~\bibnamefont{Gubitosi}},
  \bibinfo{journal}{JCAP} \textbf{\bibinfo{volume}{1503}}, \bibinfo{pages}{028}
  (\bibinfo{year}{2015}{\natexlab{b}}), \eprint{1411.7621}.

\bibitem[{\citenamefont{Bekenstein}(1993)}]{Bekenstein:1992pj}
\bibinfo{author}{\bibfnamefont{J.~D.} \bibnamefont{Bekenstein}},
  \bibinfo{journal}{Phys.Rev.} \textbf{\bibinfo{volume}{D48}},
  \bibinfo{pages}{3641} (\bibinfo{year}{1993}), \eprint{gr-qc/9211017}.

\bibitem[{\citenamefont{Khachatryan et~al.}(2014{\natexlab{a}})}]{cmsjets}
\bibinfo{author}{\bibfnamefont{V.}~\bibnamefont{Khachatryan}}
  \bibnamefont{et~al.} (\bibinfo{collaboration}{CMS})
  (\bibinfo{year}{2014}{\natexlab{a}}), \eprint{1408.3583}.

\bibitem[{\citenamefont{Khachatryan et~al.}(2014{\natexlab{b}})}]{cmsphoton}
\bibinfo{author}{\bibfnamefont{V.}~\bibnamefont{Khachatryan}}
  \bibnamefont{et~al.} (\bibinfo{collaboration}{CMS})
  (\bibinfo{year}{2014}{\natexlab{b}}), \eprint{1410.8812}.

\bibitem[{\citenamefont{Aad et~al.}(2015)}]{atlasphoton}
\bibinfo{author}{\bibfnamefont{G.}~\bibnamefont{Aad}} \bibnamefont{et~al.}
  (\bibinfo{collaboration}{ATLAS}), \bibinfo{journal}{Phys.Rev.}
  \textbf{\bibinfo{volume}{D91}}, \bibinfo{pages}{012008}
  (\bibinfo{year}{2015}), \eprint{1411.1559}.

\bibitem[{\citenamefont{Aad et~al.}(2014{\natexlab{a}})}]{atlasdilepton}
\bibinfo{author}{\bibfnamefont{G.}~\bibnamefont{Aad}} \bibnamefont{et~al.}
  (\bibinfo{collaboration}{ATLAS}), \bibinfo{journal}{Phys.Rev.}
  \textbf{\bibinfo{volume}{D90}}, \bibinfo{pages}{012004}
  (\bibinfo{year}{2014}{\natexlab{a}}), \eprint{1404.0051}.

\bibitem[{\citenamefont{Peskin and Takeuchi}(1992)}]{Peskin:1991sw}
\bibinfo{author}{\bibfnamefont{M.~E.} \bibnamefont{Peskin}} \bibnamefont{and}
  \bibinfo{author}{\bibfnamefont{T.}~\bibnamefont{Takeuchi}},
  \bibinfo{journal}{Phys.Rev.} \textbf{\bibinfo{volume}{D46}},
  \bibinfo{pages}{381} (\bibinfo{year}{1992}).

\bibitem[{\citenamefont{Peskin and Takeuchi}(1990)}]{Peskin:1990zt}
\bibinfo{author}{\bibfnamefont{M.~E.} \bibnamefont{Peskin}} \bibnamefont{and}
  \bibinfo{author}{\bibfnamefont{T.}~\bibnamefont{Takeuchi}},
  \bibinfo{journal}{Phys.Rev.Lett.} \textbf{\bibinfo{volume}{65}},
  \bibinfo{pages}{964} (\bibinfo{year}{1990}).

\bibitem[{\citenamefont{Olive et~al.}(2014)}]{Agashe:2014kda}
\bibinfo{author}{\bibfnamefont{K.}~\bibnamefont{Olive}} \bibnamefont{et~al.}
  (\bibinfo{collaboration}{Particle Data Group}), \bibinfo{journal}{Chin.Phys.}
  \textbf{\bibinfo{volume}{C38}}, \bibinfo{pages}{090001}
  (\bibinfo{year}{2014}).

\bibitem[{\citenamefont{Passarino and Veltman}(1979)}]{Passarino:1978jh}
\bibinfo{author}{\bibfnamefont{G.}~\bibnamefont{Passarino}} \bibnamefont{and}
  \bibinfo{author}{\bibfnamefont{M.}~\bibnamefont{Veltman}},
  \bibinfo{journal}{Nucl.Phys.} \textbf{\bibinfo{volume}{B160}},
  \bibinfo{pages}{151} (\bibinfo{year}{1979}).

\bibitem[{\citenamefont{Denner}(1993)}]{Denner:1991kt}
\bibinfo{author}{\bibfnamefont{A.}~\bibnamefont{Denner}},
  \bibinfo{journal}{Fortsch.Phys.} \textbf{\bibinfo{volume}{41}},
  \bibinfo{pages}{307} (\bibinfo{year}{1993}), \eprint{0709.1075}.

\bibitem[{\citenamefont{Barbieri et~al.}(2004)\citenamefont{Barbieri, Pomarol,
  Rattazzi, and Strumia}}]{Barbieri:2004qk}
\bibinfo{author}{\bibfnamefont{R.}~\bibnamefont{Barbieri}},
  \bibinfo{author}{\bibfnamefont{A.}~\bibnamefont{Pomarol}},
  \bibinfo{author}{\bibfnamefont{R.}~\bibnamefont{Rattazzi}}, \bibnamefont{and}
  \bibinfo{author}{\bibfnamefont{A.}~\bibnamefont{Strumia}},
  \bibinfo{journal}{Nucl.Phys.} \textbf{\bibinfo{volume}{B703}},
  \bibinfo{pages}{127} (\bibinfo{year}{2004}), \eprint{hep-ph/0405040}.

\bibitem[{\citenamefont{Khachatryan et~al.}(2014{\natexlab{c}})}]{cmslepton}
\bibinfo{author}{\bibfnamefont{V.}~\bibnamefont{Khachatryan}}
  \bibnamefont{et~al.} (\bibinfo{collaboration}{CMS})
  (\bibinfo{year}{2014}{\natexlab{c}}), \eprint{1408.2745}.

\bibitem[{\citenamefont{Aad et~al.}(2014{\natexlab{b}})}]{atlaslepton}
\bibinfo{author}{\bibfnamefont{G.}~\bibnamefont{Aad}} \bibnamefont{et~al.}
  (\bibinfo{collaboration}{ATLAS}), \bibinfo{journal}{Phys.Rev.}
  \textbf{\bibinfo{volume}{D90}}, \bibinfo{pages}{012004}
  (\bibinfo{year}{2014}{\natexlab{b}}), \eprint{1404.0051}.

\bibitem[{\citenamefont{Alloul et~al.}(2014)\citenamefont{Alloul, Christensen,
  Degrande, Duhr, and Fuks}}]{feynrules}
\bibinfo{author}{\bibfnamefont{A.}~\bibnamefont{Alloul}},
  \bibinfo{author}{\bibfnamefont{N.~D.} \bibnamefont{Christensen}},
  \bibinfo{author}{\bibfnamefont{C.}~\bibnamefont{Degrande}},
  \bibinfo{author}{\bibfnamefont{C.}~\bibnamefont{Duhr}}, \bibnamefont{and}
  \bibinfo{author}{\bibfnamefont{B.}~\bibnamefont{Fuks}},
  \bibinfo{journal}{Comput.Phys.Commun.} \textbf{\bibinfo{volume}{185}},
  \bibinfo{pages}{2250} (\bibinfo{year}{2014}), \eprint{1310.1921}.

\bibitem[{\citenamefont{Alwall et~al.}(2011)\citenamefont{Alwall, Herquet,
  Maltoni, Mattelaer, and Stelzer}}]{madevent}
\bibinfo{author}{\bibfnamefont{J.}~\bibnamefont{Alwall}},
  \bibinfo{author}{\bibfnamefont{M.}~\bibnamefont{Herquet}},
  \bibinfo{author}{\bibfnamefont{F.}~\bibnamefont{Maltoni}},
  \bibinfo{author}{\bibfnamefont{O.}~\bibnamefont{Mattelaer}},
  \bibnamefont{and} \bibinfo{author}{\bibfnamefont{T.}~\bibnamefont{Stelzer}},
  \bibinfo{journal}{JHEP} \textbf{\bibinfo{volume}{1106}}, \bibinfo{pages}{128}
  (\bibinfo{year}{2011}), \eprint{1106.0522}.

\bibitem[{\citenamefont{Gleisberg et~al.}(2009)\citenamefont{Gleisberg, Hoeche,
  Krauss, Schonherr, Schumann et~al.}}]{sherpa}
\bibinfo{author}{\bibfnamefont{T.}~\bibnamefont{Gleisberg}},
  \bibinfo{author}{\bibfnamefont{S.}~\bibnamefont{Hoeche}},
  \bibinfo{author}{\bibfnamefont{F.}~\bibnamefont{Krauss}},
  \bibinfo{author}{\bibfnamefont{M.}~\bibnamefont{Schonherr}},
  \bibinfo{author}{\bibfnamefont{S.}~\bibnamefont{Schumann}},
  \bibnamefont{et~al.}, \bibinfo{journal}{JHEP}
  \textbf{\bibinfo{volume}{0902}}, \bibinfo{pages}{007} (\bibinfo{year}{2009}),
  \eprint{0811.4622}.

\bibitem[{\citenamefont{Bahr et~al.}(2008)\citenamefont{Bahr, Gieseke, Gigg,
  Grellscheid, Hamilton et~al.}}]{herwig}
\bibinfo{author}{\bibfnamefont{M.}~\bibnamefont{Bahr}},
  \bibinfo{author}{\bibfnamefont{S.}~\bibnamefont{Gieseke}},
  \bibinfo{author}{\bibfnamefont{M.}~\bibnamefont{Gigg}},
  \bibinfo{author}{\bibfnamefont{D.}~\bibnamefont{Grellscheid}},
  \bibinfo{author}{\bibfnamefont{K.}~\bibnamefont{Hamilton}},
  \bibnamefont{et~al.}, \bibinfo{journal}{Eur.Phys.J.}
  \textbf{\bibinfo{volume}{C58}}, \bibinfo{pages}{639} (\bibinfo{year}{2008}),
  \eprint{0803.0883}.

\bibitem[{\citenamefont{Cacciari et~al.}(2012)\citenamefont{Cacciari, Salam,
  and Soyez}}]{fastjet}
\bibinfo{author}{\bibfnamefont{M.}~\bibnamefont{Cacciari}},
  \bibinfo{author}{\bibfnamefont{G.~P.} \bibnamefont{Salam}}, \bibnamefont{and}
  \bibinfo{author}{\bibfnamefont{G.}~\bibnamefont{Soyez}},
  \bibinfo{journal}{Eur.Phys.J.} \textbf{\bibinfo{volume}{C72}},
  \bibinfo{pages}{1896} (\bibinfo{year}{2012}), \eprint{1111.6097}.

\bibitem[{\citenamefont{Englert and Spannowsky}(2015)}]{Englert:2014cva}
\bibinfo{author}{\bibfnamefont{C.}~\bibnamefont{Englert}} \bibnamefont{and}
  \bibinfo{author}{\bibfnamefont{M.}~\bibnamefont{Spannowsky}},
  \bibinfo{journal}{Phys.Lett.} \textbf{\bibinfo{volume}{B740}},
  \bibinfo{pages}{8} (\bibinfo{year}{2015}), \eprint{1408.5147}.

\bibitem[{\citenamefont{Buchmueller et~al.}(2014)\citenamefont{Buchmueller,
  Dolan, and McCabe}}]{Buchmueller:2013dya}
\bibinfo{author}{\bibfnamefont{O.}~\bibnamefont{Buchmueller}},
  \bibinfo{author}{\bibfnamefont{M.~J.} \bibnamefont{Dolan}}, \bibnamefont{and}
  \bibinfo{author}{\bibfnamefont{C.}~\bibnamefont{McCabe}},
  \bibinfo{journal}{JHEP} \textbf{\bibinfo{volume}{1401}}, \bibinfo{pages}{025}
  (\bibinfo{year}{2014}), \eprint{1308.6799}.

\bibitem[{\citenamefont{Harris et~al.}(2015)\citenamefont{Harris, Khoze,
  Spannowsky, and Williams}}]{Harris:2014hga}
\bibinfo{author}{\bibfnamefont{P.}~\bibnamefont{Harris}},
  \bibinfo{author}{\bibfnamefont{V.~V.} \bibnamefont{Khoze}},
  \bibinfo{author}{\bibfnamefont{M.}~\bibnamefont{Spannowsky}},
  \bibnamefont{and} \bibinfo{author}{\bibfnamefont{C.}~\bibnamefont{Williams}},
  \bibinfo{journal}{Phys.Rev.} \textbf{\bibinfo{volume}{D91}},
  \bibinfo{pages}{055009} (\bibinfo{year}{2015}), \eprint{1411.0535}.

\bibitem[{\citenamefont{Jacques and Nordström}(2015)}]{Jacques:2015zha}
\bibinfo{author}{\bibfnamefont{T.}~\bibnamefont{Jacques}} \bibnamefont{and}
  \bibinfo{author}{\bibfnamefont{K.}~\bibnamefont{Nordström}}
  (\bibinfo{year}{2015}), \eprint{1502.05721}.

\bibitem[{\citenamefont{Buckley et~al.}(2015)\citenamefont{Buckley, Feld, and
  Goncalves}}]{Buckley:2014fba}
\bibinfo{author}{\bibfnamefont{M.~R.} \bibnamefont{Buckley}},
  \bibinfo{author}{\bibfnamefont{D.}~\bibnamefont{Feld}}, \bibnamefont{and}
  \bibinfo{author}{\bibfnamefont{D.}~\bibnamefont{Goncalves}},
  \bibinfo{journal}{Phys.Rev.} \textbf{\bibinfo{volume}{D91}},
  \bibinfo{pages}{015017} (\bibinfo{year}{2015}), \eprint{1410.6497}.

\bibitem[{\citenamefont{Haisch and Re}(2015)}]{Haisch:2015ioa}
\bibinfo{author}{\bibfnamefont{U.}~\bibnamefont{Haisch}} \bibnamefont{and}
  \bibinfo{author}{\bibfnamefont{E.}~\bibnamefont{Re}} (\bibinfo{year}{2015}),
  \eprint{1503.00691}.

\bibitem[{\citenamefont{Isidori et~al.}(2014)\citenamefont{Isidori, Manohar,
  and Trott}}]{Isidori:2013cla}
\bibinfo{author}{\bibfnamefont{G.}~\bibnamefont{Isidori}},
  \bibinfo{author}{\bibfnamefont{A.~V.} \bibnamefont{Manohar}},
  \bibnamefont{and} \bibinfo{author}{\bibfnamefont{M.}~\bibnamefont{Trott}},
  \bibinfo{journal}{Phys.Lett.} \textbf{\bibinfo{volume}{B728}},
  \bibinfo{pages}{131} (\bibinfo{year}{2014}), \eprint{1305.0663}.

\bibitem[{\citenamefont{Cacciari et~al.}(2008)\citenamefont{Cacciari, Salam,
  and Soyez}}]{antikt}
\bibinfo{author}{\bibfnamefont{M.}~\bibnamefont{Cacciari}},
  \bibinfo{author}{\bibfnamefont{G.~P.} \bibnamefont{Salam}}, \bibnamefont{and}
  \bibinfo{author}{\bibfnamefont{G.}~\bibnamefont{Soyez}},
  \bibinfo{journal}{JHEP} \textbf{\bibinfo{volume}{0804}}, \bibinfo{pages}{063}
  (\bibinfo{year}{2008}), \eprint{0802.1189}.

\bibitem[{\citenamefont{Salam}(2010)}]{Salam:2009jx}
\bibinfo{author}{\bibfnamefont{G.~P.} \bibnamefont{Salam}},
  \bibinfo{journal}{Eur.Phys.J.} \textbf{\bibinfo{volume}{C67}},
  \bibinfo{pages}{637} (\bibinfo{year}{2010}), \eprint{0906.1833}.

\bibitem[{\citenamefont{Englert et~al.}(2012)\citenamefont{Englert, Spannowsky,
  and Wymant}}]{Englert:2012wf}
\bibinfo{author}{\bibfnamefont{C.}~\bibnamefont{Englert}},
  \bibinfo{author}{\bibfnamefont{M.}~\bibnamefont{Spannowsky}},
  \bibnamefont{and} \bibinfo{author}{\bibfnamefont{C.}~\bibnamefont{Wymant}},
  \bibinfo{journal}{Phys.Lett.} \textbf{\bibinfo{volume}{B718}},
  \bibinfo{pages}{538} (\bibinfo{year}{2012}), \eprint{1209.0494}.

\bibitem[{\citenamefont{Read}(2002)}]{cls1}
\bibinfo{author}{\bibfnamefont{A.~L.} \bibnamefont{Read}},
  \bibinfo{journal}{J.Phys.} \textbf{\bibinfo{volume}{G28}},
  \bibinfo{pages}{2693} (\bibinfo{year}{2002}).

\bibitem[{\citenamefont{Read}(2000)}]{cls2}
\bibinfo{author}{\bibfnamefont{A.~L.} \bibnamefont{Read}},
  \bibinfo{journal}{CERN-OPEN-2000-205}  (\bibinfo{year}{2000}).

\bibitem[{\citenamefont{Englert et~al.}(2014)\citenamefont{Englert, Freitas,
  Mühlleitner, Plehn, Rauch et~al.}}]{Englert:2014uua}
\bibinfo{author}{\bibfnamefont{C.}~\bibnamefont{Englert}},
  \bibinfo{author}{\bibfnamefont{A.}~\bibnamefont{Freitas}},
  \bibinfo{author}{\bibfnamefont{M.}~\bibnamefont{Mühlleitner}},
  \bibinfo{author}{\bibfnamefont{T.}~\bibnamefont{Plehn}},
  \bibinfo{author}{\bibfnamefont{M.}~\bibnamefont{Rauch}},
  \bibnamefont{et~al.}, \bibinfo{journal}{J.Phys.}
  \textbf{\bibinfo{volume}{G41}}, \bibinfo{pages}{113001}
  (\bibinfo{year}{2014}), \eprint{1403.7191}.

\bibitem[{\citenamefont{Klute et~al.}(2013)\citenamefont{Klute, Lafaye, Plehn,
  Rauch, and Zerwas}}]{Klute:2013cx}
\bibinfo{author}{\bibfnamefont{M.}~\bibnamefont{Klute}},
  \bibinfo{author}{\bibfnamefont{R.}~\bibnamefont{Lafaye}},
  \bibinfo{author}{\bibfnamefont{T.}~\bibnamefont{Plehn}},
  \bibinfo{author}{\bibfnamefont{M.}~\bibnamefont{Rauch}}, \bibnamefont{and}
  \bibinfo{author}{\bibfnamefont{D.}~\bibnamefont{Zerwas}},
  \bibinfo{journal}{Europhys.Lett.} \textbf{\bibinfo{volume}{101}},
  \bibinfo{pages}{51001} (\bibinfo{year}{2013}), \eprint{1301.1322}.

\bibitem[{\citenamefont{Kniehl and Spira}(1995)}]{Kniehl:1995tn}
\bibinfo{author}{\bibfnamefont{B.~A.} \bibnamefont{Kniehl}} \bibnamefont{and}
  \bibinfo{author}{\bibfnamefont{M.}~\bibnamefont{Spira}},
  \bibinfo{journal}{Z.Phys.} \textbf{\bibinfo{volume}{C69}},
  \bibinfo{pages}{77} (\bibinfo{year}{1995}), \eprint{hep-ph/9505225}.

\bibitem[{\citenamefont{Ellis et~al.}(1976)\citenamefont{Ellis, Gaillard, and
  Nanopoulos}}]{Ellis:1975ap}
\bibinfo{author}{\bibfnamefont{J.~R.} \bibnamefont{Ellis}},
  \bibinfo{author}{\bibfnamefont{M.~K.} \bibnamefont{Gaillard}},
  \bibnamefont{and} \bibinfo{author}{\bibfnamefont{D.~V.}
  \bibnamefont{Nanopoulos}}, \bibinfo{journal}{Nucl.Phys.}
  \textbf{\bibinfo{volume}{B106}}, \bibinfo{pages}{292} (\bibinfo{year}{1976}).

\bibitem[{\citenamefont{Shifman et~al.}(1979)\citenamefont{Shifman, Vainshtein,
  Voloshin, and Zakharov}}]{Shifman:1979eb}
\bibinfo{author}{\bibfnamefont{M.~A.} \bibnamefont{Shifman}},
  \bibinfo{author}{\bibfnamefont{A.}~\bibnamefont{Vainshtein}},
  \bibinfo{author}{\bibfnamefont{M.}~\bibnamefont{Voloshin}}, \bibnamefont{and}
  \bibinfo{author}{\bibfnamefont{V.~I.} \bibnamefont{Zakharov}},
  \bibinfo{journal}{Sov.J.Nucl.Phys.} \textbf{\bibinfo{volume}{30}},
  \bibinfo{pages}{711} (\bibinfo{year}{1979}).

\bibitem[{\citenamefont{Buchalla et~al.}(1996)\citenamefont{Buchalla, Buras,
  and Lautenbacher}}]{Buchalla:1995vs}
\bibinfo{author}{\bibfnamefont{G.}~\bibnamefont{Buchalla}},
  \bibinfo{author}{\bibfnamefont{A.~J.} \bibnamefont{Buras}}, \bibnamefont{and}
  \bibinfo{author}{\bibfnamefont{M.~E.} \bibnamefont{Lautenbacher}},
  \bibinfo{journal}{Rev.Mod.Phys.} \textbf{\bibinfo{volume}{68}},
  \bibinfo{pages}{1125} (\bibinfo{year}{1996}), \eprint{hep-ph/9512380}.

\bibitem[{\citenamefont{Hui and Nicolis}(2010)}]{Hui:2010dn}
\bibinfo{author}{\bibfnamefont{L.}~\bibnamefont{Hui}} \bibnamefont{and}
  \bibinfo{author}{\bibfnamefont{A.}~\bibnamefont{Nicolis}},
  \bibinfo{journal}{Phys.Rev.Lett.} \textbf{\bibinfo{volume}{105}},
  \bibinfo{pages}{231101} (\bibinfo{year}{2010}), \eprint{1009.2520}.

\bibitem[{\citenamefont{Englert et~al.}(2013)\citenamefont{Englert, Jaeckel,
  Khoze, and Spannowsky}}]{Englert:2013gz}
\bibinfo{author}{\bibfnamefont{C.}~\bibnamefont{Englert}},
  \bibinfo{author}{\bibfnamefont{J.}~\bibnamefont{Jaeckel}},
  \bibinfo{author}{\bibfnamefont{V.}~\bibnamefont{Khoze}}, \bibnamefont{and}
  \bibinfo{author}{\bibfnamefont{M.}~\bibnamefont{Spannowsky}},
  \bibinfo{journal}{JHEP} \textbf{\bibinfo{volume}{1304}}, \bibinfo{pages}{060}
  (\bibinfo{year}{2013}), \eprint{1301.4224}.

\bibitem[{\citenamefont{Coleman and Mandula}(1967)}]{Coleman:1967ad}
\bibinfo{author}{\bibfnamefont{S.~R.} \bibnamefont{Coleman}} \bibnamefont{and}
  \bibinfo{author}{\bibfnamefont{J.}~\bibnamefont{Mandula}},
  \bibinfo{journal}{Phys.Rev.} \textbf{\bibinfo{volume}{159}},
  \bibinfo{pages}{1251} (\bibinfo{year}{1967}).

\bibitem[{\citenamefont{Brown}(1992)}]{Brown:1992db}
\bibinfo{author}{\bibfnamefont{L.}~\bibnamefont{Brown}},
  \emph{\bibinfo{title}{{Quantum field theory}}} (\bibinfo{year}{1992}).

\end{thebibliography}

\end{document}